\documentclass[preprint]{aastex}
\shorttitle{RZ2109 Abundances}
\shortauthors{Steele et al.}

\newcommand{\kms}{km\thinspace s$^{-1}$ }
\newcommand{\ovh}{[O~III]$\lambda$5007/H$\beta$ }
\newcommand{\zeroseven}{[O~III]$\lambda$5007 }

\begin{document}
\title{Composition of an Emission Line System in Black Hole Host
Globular Cluster RZ2109}

\author{Matthew M. Steele\altaffilmark{1}\altaffilmark{2}, Stephen E. Zepf\altaffilmark{1},
  Thomas J. Maccarone\altaffilmark{3}, Arunav Kundu\altaffilmark{4}, 
Katherine L. Rhode\altaffilmark{5}, and John J. Salzer\altaffilmark{5}}

\altaffiltext{1}{Department of Physics \& Astronomy, Michigan State University,
  East Lansing, MI 48824; e-mail: steele24@msu.edu}
\altaffiltext{2}{Department of Physics, Northern Michigan University, Marquette,
  MI 49855}
\altaffiltext{3}{Department of Physics, Texas Tech University, Lubbock, TX 79409}
\altaffiltext{4}{Eureka Scientific, 2452 Delmer Street Suite 100, Oakland, CA 94602-3017}
\altaffiltext{5}{Department of Astronomy, Indiana University, Bloomington, IN 47405}

\begin{abstract}
We present an analysis of optical spectra from the globular cluster RZ2109 in
NGC~4472, which hosts the first unambiguous globular cluster black hole.  We
use these spectra to determine the elemental composition of the emission line
system associated with this source, and to constrain the age and metallicity of
the host globular cluster.  For the emission line system of RZ2109, our
analysis indicates the \zeroseven equivalent width is $33.82 \pm 0.39$ \AA\ and
the H$\beta$ equivalent width is $0.32 \pm 0.32$ \AA\ , producing a formal \ovh
emission line ratio of 106 for a 3200 \kms measurement aperture covering the
full velocity width of the [O~III]$\lambda$5007 line. Within a narrower 600
\kms aperture covering the highest luminosity velocity structure in the line
complex, we find \ovh = 62.  The measured \ovh ratios are significantly higher
than can be produced in radiative models of the emission line region with solar
composition, and the confidence interval limits exclude all but models which
have gas masses much larger than those for a single star. Therefore, we
conclude that the region from which the [O~III]$\lambda$5007 emission
originates is hydrogen-depleted relative to solar composition gas. This finding
is consistent with emission from an accretion-powered outflow driven by a
hydrogen-depleted donor star, such as a white dwarf, being accreted onto a
black hole.

\end{abstract}

\section{INTRODUCTION}
The globular cluster RZ2109 located in the galaxy NGC4472 is a known host of an
accreting black hole system, first identified by \citet{Maccarone2007}.  Along
with the variable X-ray source indicative of the accreting black hole, RZ2109
has been observed to host a broad and luminous [O~III]$\lambda
\lambda$4959,5007 emission line complex thought to correspond to an
accretion-powered outflow driven from the cluster's black hole
\citep{Zepf2007,Zepf2008,Steele2011}.

The RZ2109 emission line structure is unusual in many respects. The lines are
very broad with a width of 3200 km s$^{-1}$, about a factor of 100 larger than
the escape velocity of the globular cluster in which it is located.  The line
velocity profiles also have a complex shape and appear to contain two distinct
velocity structures \citep{Steele2011}.  Moreover, [O~III]$\lambda
\lambda$4959,5007 are the only emission lines apparent even in a fairly deep
spectrum \citep{Zepf2008}. \citet{Gnedin2009} suggested that the lack of Balmer
line emission, in particular, may be indicative of a hydrogen poor donor star
such as a white dwarf. This possibility is significant because although BH-WD
binaries are an expected end stage of stellar evolution, especially in globular
clusters, none has yet been positively identified \citep{Ivanova2011}.

A common diagnostic for the study of emission line regions is the \ovh emission
line ratio. Previous work by \citet{Zepf2008} noted that \ovh ratio appeared to
be at least 30 based on a low resolution optical spectrum of the RZ2109
cluster. For reference other common astrophysical [O~III] emission line
production sites include active galactic nuclei with typical \ovh ratios of
order unity \citep{Sarzi2006}, Milky Way planetary nebulae with mean line
ratios of approximately 15 \citep{Stanghellini2003}, and a few extremely
hydrogen poor planetary nebulae with line ratio of ~20
\citep{Mendez2005,Larsen2008}.  Based on a comparison to these sources, it
seems plausible that the RZ2109 emission line site would be hydrogen-depleted.
On the other hand, the [O III] production sites and mechanisms can be very
different among astrophysical object classes and it is not immediately obvious
how relevant such comparisons are.

In this work we investigate the composition of the RZ2109 emission line region
in order to help constrain models of the black hole/donor star binary in the
globular cluster.  Accordingly, Section \ref{sec:obs} presents an analysis of
higher signal-to-noise spectra of RZ2109 than have been previously used to
explore the composition of the source.  In Section \ref{sec:ar}, we begin by
modeling the stellar component of the RZ2109 spectrum to remove the stellar
contributions from the emission line measurements.  The next section presents
the line measurement results.  Section \ref{sec:rtm} presents results from a
suite of radiative models that we use to interpret the range of \ovh emission
line ratios.  The last section of the paper includes a discussion of the
overall results of this analysis.

\section{OBSERVATIONS AND DATA REDUCTION} \label{sec:obs}
The optical spectra used in this study were collected using the Gemini Multi
Object Spectrograph (GMOS) on the Gemini South Telescope.  These data, first
presented in \citet{Steele2011}, were taken on the consecutive nights of UT 2009
March 28--29 (program GS-2009A-Q-1). The medium resolution spectra (R $\sim
2400$) have a wavelength coverage from 4445--6023 \AA.  The data were reduced
using the longslit tools in the Gemini IRAF package. One dimensional spectra
were extracted from the processed spectra and, following tests to ensure the
emission line object was not varying between nights \citep{Steele2011},
co-added to increase the signal-to-noise ratio.

The resulting reduced spectrum is a superposition of the emission line source of
interest and the stellar component of the host globular cluster. In order to
examine the emission line source in isolation, the stellar component must be
modeled and removed.

\section{ANALYSIS AND RESULTS} \label{sec:ar}
\subsection{Stellar Component Modeling} \label{sec:sm}

The modeling of the stellar component was performed in a process similar to
that described in \citet{Steele2011}. The RZ2109 Gemini spectrum was fit to a
grid of continuum normalized synthetic spectra constructed using the
Pegas\'e-HR code of \citet{LeBorgne2004} and Elodie 3.1 \citep{Prugniel2007}
stellar library. A Kroupa IMF \citep{Kroupa2001} was adopted along with a
single epoch of star formation to construct the synthetic stellar population
grid using cluster age and initial metallicity as free parameters. To avoid
emission lines contaminating the fit with the inclusion of emission lines with
non-stellar origins, the region between 4870--5060 \AA\ in the observed frame
(4846--5035 \AA\ rest frame)
was masked out to exclude contributions from H$\beta$ and [O~III]$\lambda$5007
to the $\chi^2$ minimization fitting scheme. The resulting best-fit stellar
model corresponded to an age of 13.25 Gyr and a metallicity of [Fe/H] = -1.0.

\begin{figure}
  \begin{center}
    \includegraphics[scale=0.8]{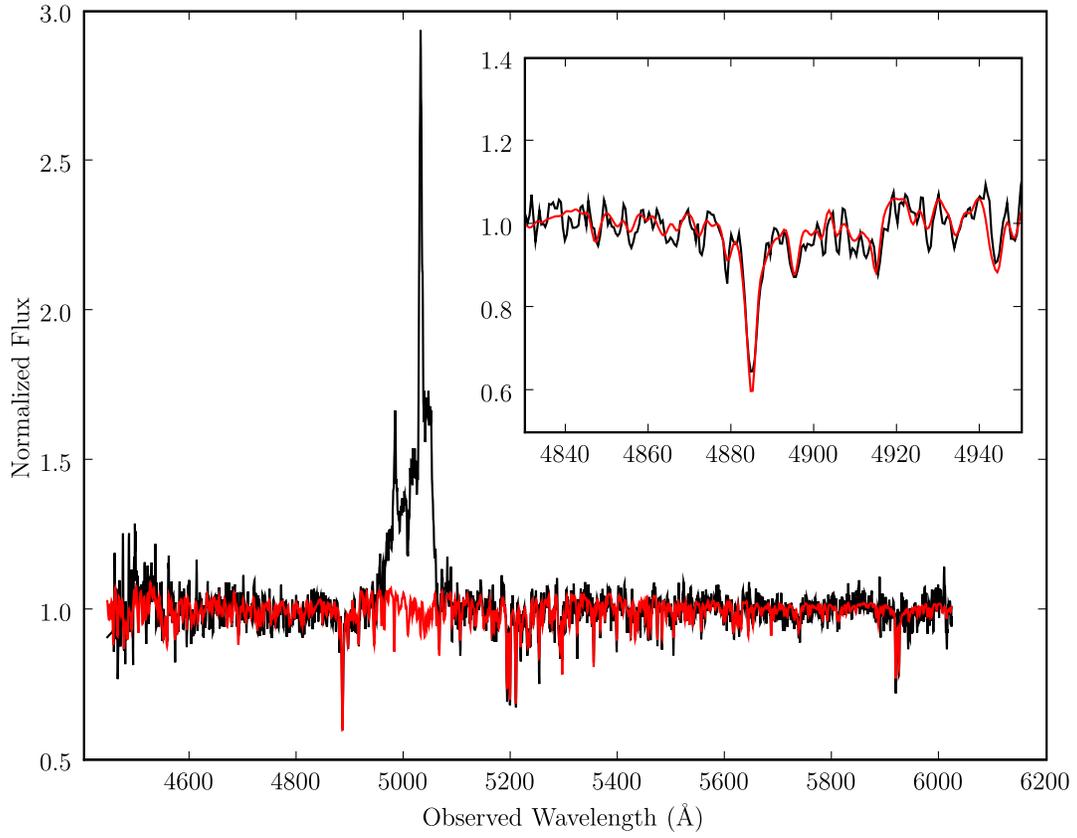}
    \caption{GMOS spectrum and the synthetic stellar component model. The GMOS
      observations of RZ2109 are shown using the black line. The gray line (red
      in the electronic version) depicts the best-fit synthetic stellar model.
      The inset shows the wavelength region surrounding H$\beta$ in
      detail.} \label{fig:stellar}
  \end{center}
\end{figure}

The best-fitting synthetic stellar population model provides, qualitatively, an
excellent match to the stellar component of the observed spectrum of
RZ2109. Figure \ref{fig:stellar} provides a comparison between the observed
spectrum and the best-fit model over the entire spectral wavelength range,
along with an inset showing the H$\beta$ region in detail.

\begin{figure}
  \begin{center}
    \includegraphics[scale=0.8]{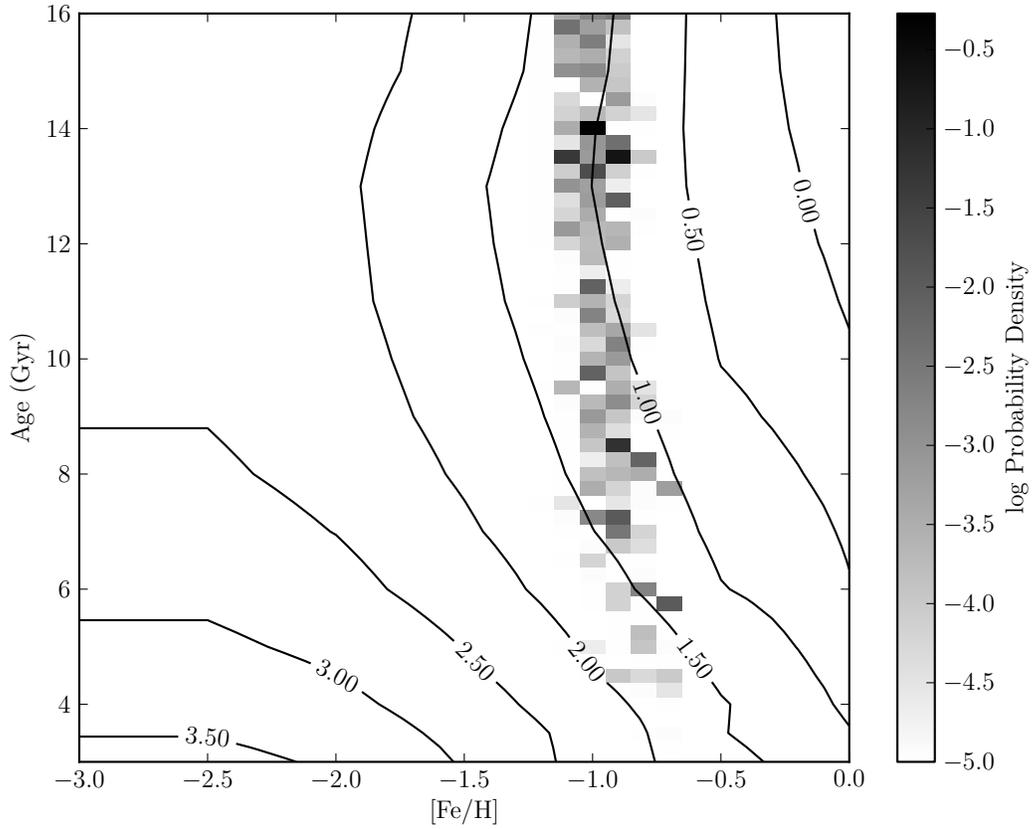}
    \caption{Stellar Model Fitting Uncertainties and H$\beta$ equivalent width
      contours. The gray scale grid displays the probability density in stellar
      parameter space that the stellar component of the RZ2109 spectrum would be
      determined to have a given set of parameters as described in section
      \ref{sec:sm}. The contour lines show the measurement of the H$\beta$
      equivalent width with a 600 \kms aperture for a given stellar component
      model relative to the best-fit model, as described in section
      \ref{sec:ew}.} \label{fig:sig}
  \end{center}
\end{figure}

In order to determine the uncertainties in the fitted parameters, simulated
observations were performed by constructing a best-fit synthetic stellar
component spectrum and injecting noise to match the signal-to-noise of the
observations. The simulated observation was then fit to produce a new set of
stellar parameters. This simulated observation process was performed for 10$^5$
iterations with the resultant probability density distribution of the stellar
component parameter fits shown in Figure \ref{fig:sig}.  From this exercise it
is evident that the fit is well constrained in [Fe/H], while displaying a
greater spread in the age parameter. Some of the observed spread in age results
from the exclusion of H$\beta$, the feature most sensitive to age, in the
fitting scheme. The log scale used for the stellar parameter density
distribution in Figure \ref{fig:sig} may overemphasize the spread in age; only
1 percent of the simulated observations had age determinations below 5.75 Gyr
and only 10 percent were below 8.50 Gyr. The adopted stellar parameters and
associated uncertainties (Age $=13.25 \pm 1.00$ Gyr, [Fe/H] $=-1.0 \pm 0.1$)
include 83 percent of all simulated observations in age and 98 percent in
metallicity.

\subsection{Equivalent Widths} \label{sec:ew}

\begin{figure}
  \begin{center}
    \includegraphics[scale=0.8]{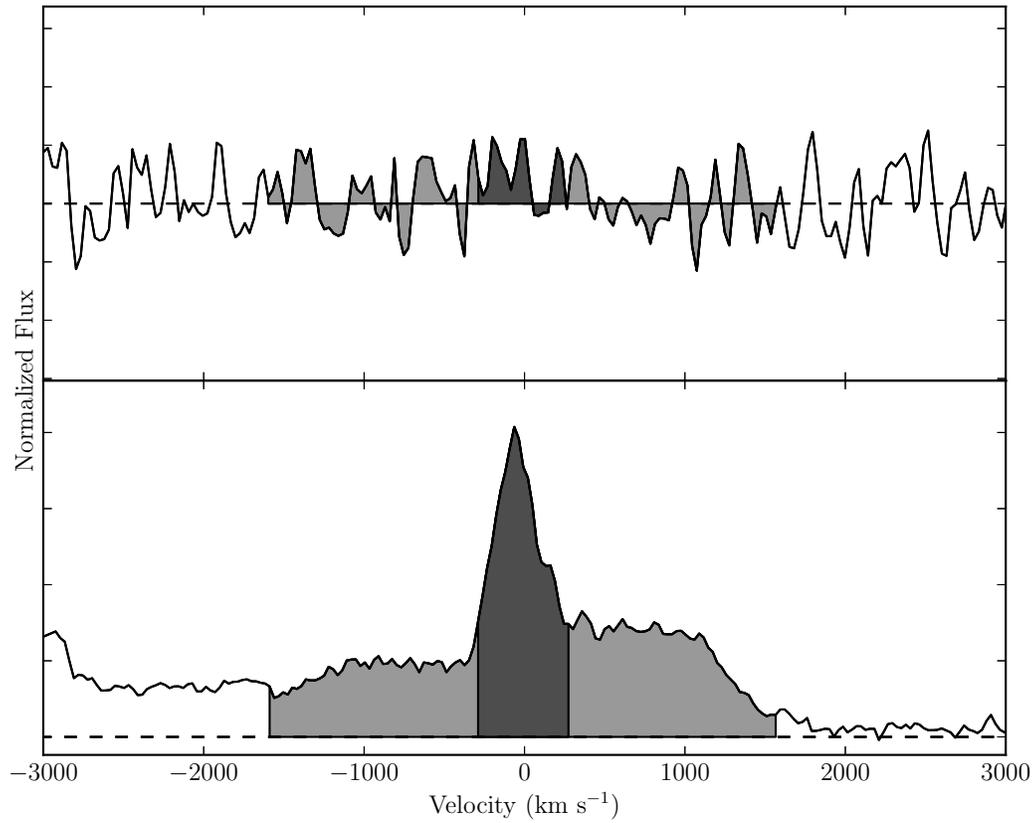}
    \caption{Equivalent Width apertures. The upper panel displays the H$\beta$
      region of the stellar model subtracted GMOS spectrum in velocity space,
      and the lower panel gives the [O~III]$\lambda$5007 region. The 600 \kms
      aperture over which the equivalent width was measured is shaded with dark
      gray. The 3200 \kms aperture is shown in light gray. The dashed line
      gives the continuum level.}\label{fig:aps}
  \end{center}
\end{figure}

The emission line's equivalent widths (EW) are measured by direct integration of
the continuum normalized spectrum bounded by a given velocity aperture. A second
measurement is then performed over the identical aperture on the stellar model
and the value subtracted from the data measurement in order to remove the
influence of stellar absorption features. Given the complexity of the
[O~III]$\lambda$5007 line profile the specification of the velocity aperture is
non-trivial. \citet{Steele2011} find that the [O~III]$\lambda$5007 emission line
in RZ2109 is well described by two components; a broad component with a width at
zero intensity of $\sim 3200$ \kms and narrower component with a FWHM of $\sim
300$ \kms. Here we perform measurements for apertures of two widths which
correspond to the two components identified by \citet{Steele2011}; a broad 3200
\kms aperture and a 600 \kms aperture. For reference the apertures over which
the measurements are performed are plotted for the H$\beta$ and
[O~III]$\lambda$5007 lines in figure \ref{fig:aps}. Table \ref{tab:ew} gives the
resulting measurements for [O~III]$\lambda$5007 and H$\beta$ in both the 3200
and 600 \kms apertures. The \ovh emission line ratios for each velocity aperture
are very large (105.7 for the 3200 \kms aperture, 61.6 for the 600 \kms
aperture), notably larger than found in previous measurements using lower
signal-to-noise data \citep{Zepf2008}. This increase in the emission line ratios
may be attributable solely to the higher signal-to-noise of the current data
providing lower H$\beta$ measurement uncertainties, and therefore higher line
ratios, rather than a change in the line luminosities of the source.

The cited 1$\sigma$ uncertainties provide an estimate of the RMS uncertainties
derived from the local continuum and Poisson statistic uncertainties for
emission above the local continuum. To construct an estimate of the local
continuum RMS two regions, one blueward, one redward, with a width equal to that
of the measurement aperture were evaluated to find the mean expected deviations
from the local continuum. The second uncertainty cited for the 600 \kms aperture
H$\beta$ measurement reflects the uncertainty produced by the fit of the
synthetic stellar population model. This uncertainty estimate was produced by
injecting noise consistent with that of the observations into a synthetic
stellar model with parameters matching the best-fit model, refitting the
resulting spectrum with a new stellar model using the procedure described in
Section \ref{sec:sm}, and remeasuring the H$\beta$ EW.

The contours plotted in Figure \ref{fig:sig} display the H$\beta$ EW as a
function of the stellar component parameters. The unit label for each contour
is given as a multiple of the 600 \kms aperture EW measurement of the best-fit
stellar parameter model. Considering the full stellar parameter space covered
by the simulated observations detailed in Section \ref{sec:sm}, the mean
H$\beta$ EW of the simulated observations is two percent greater than the cited
value for the 600 \kms aperture of $0.21$ \AA, with 58 percent of the
iterations falling within 10 percent and 89 percent of iterations within 20
percent of the cited value.  If we limit consideration to only those simulated
observations located within the uncertainty bounds of the best-fit stellar
parameters, the mean H$\beta$ EW is one percent less than the cited value, 66
percent of iterations are within 10 percent of the cited value and all
simulated observations are within 20 percent.  Therefore an uncertainty in the
600 \kms aperture H$\beta$ EW measurement of 20 percent is adopted.

The uncertainty estimates of the \ovh emission line ratios are dominated by
uncertainties in the H$\beta$ measurement.  For small values of
H$\beta$ the discontinuity in the \ovh ratios, \ovh goes to infinity as H$\beta$
goes to zero, complicates the interpretation of uncertainties expressed in the
conventional format.  For example, the 3200 \kms
aperture emission line ratio is 105.7$^{+\infty}_{-52.9}$. For clarity it is
helpful to consider the uncertainties of the \ovh ratio in terms of limits at
set confidence intervals, as expressed in Figure \ref{tab:lim}. Further
discussion of these limits and there interpretation is included in Section
\ref{sec:dc}.

It should be noted that ratios of emission line EW are not always directly
comparable to ratios of emission line fluxes. EWs are measured relative to the
local continuum level, thus two emission lines with equal flux values will only
have equal EW values if the continuum level is the same at each lines observed
wavelength. In the case of RZ2109 measurements of the flux calibrated continuum
at the observed line-center of \zeroseven and H$\beta$ differ by $\sim 1$
percent, indicating that the \ovh EW ratio is functionally equivalent (within
measurement uncertainties) to a flux ratio.

Additional measurements were performed for a selection of other emission lines
which the synthetic spectral modeling of Section \ref{sec:rtm} suggested may
be present with EW a significant fraction of that of H$\beta$. These
measurements were performed with a 600 \kms aperture on He~II$\lambda$4686
(EW$=0.02 \pm 0.12$), [Ar~IV]$\lambda$4740 (EW$=0.07 \pm 0.13$), and
[Fe~VII]$\lambda$5721 (EW$=0.07 \pm 0.08$).  The uncertainties for these
measurements reflect 1 sigma RMS deviations from the continuum, determined as
describe above for the [O~III]$\lambda$5007 and H$\beta$ measurements. It
should be noted that other than [O~III]$\lambda$5007 and the 600 \kms aperture
H$\beta$ EW (which is a very weak detection), all EW for both 3200 and 600 \kms
apertures are consistent with non-detections.

As the two emission components are thought to be produced in spatially and
geometrically distinct structures, the emission line ratios derived using the
two apertures need not be the same. Since the contributions of the broad and
narrow emission components are superimposed, both the 3200 \kms or 600 \kms
aperture measurements reflect an EW with contributions from both sources. In
the 3200 \kms aperture the broad component dominates the EW, while the narrow
component dominates in the 600 \kms aperture measurements. For reference,
\citet{Steele2011} find that 81 percent of the total [O~III]$\lambda$5007 flux
is contributed by the broad component and the remainder by the narrow
component, though it should be noted deconvolving these contribution is
necessarily a model-dependent calculation.

\subsection{Emission Line System Modeling} \label{sec:rtm}

In order to test the possibility that the measured \ovh
emission line ratios given above might be produced by a gas of solar
composition, a series of radiative transfer models was constructed. The models
were executed using version 08.01 of the Cloudy spectral synthesis code from
\citet{Ferland1998}.

The suite of models tested all had solar compositions and spherically symmetric
shell gas distributions characterized by an inner and outer radius . The gas was
assumed to be outflowing at a velocity of 300 \kms at the inner radius. The gas
was gradually accelerated by radiation pressure from the X-ray source,
increasing the velocity by a few percent at the outer radius. Under the
prescription of an outflowing gas Cloudy adjusts the density to conserve mass
flux (defined as the quantity $\rho(r) r^2 u(r)$, where $\rho$ is the gas
density, $u$ is the velocity, and $r$ the radius) at each radius. The gas was
irradiated by bremsstrahlung emission from a central source at 10$^6$ K, and
normalized to match the X-ray luminosity observed by \citet{Shih2008} and
\citet{Maccarone2007}.  Parameter space was then explored along the dimensions
of the the inner radius ($R_0$), hydrogen density at the inner radius
($\rho_H(R_0)$), and the total mass of the outflow. The free parameter space
included in the model grid is summarized in Table \ref{tab:cmg}.  For each
outflow mass range the limits on the inner radius and hydrogen density were
selected to ensure the maximum \ovh ratio model was surround by closed emission
line ratio contours in parameter space. This criteria ensured that the maximum
line ratio model identified at each outflow mass was a maximum in the local
parameter space and not limited by the selection of boundary conditions.

\begin{figure}
  \begin{center}
    \includegraphics[scale=0.8]{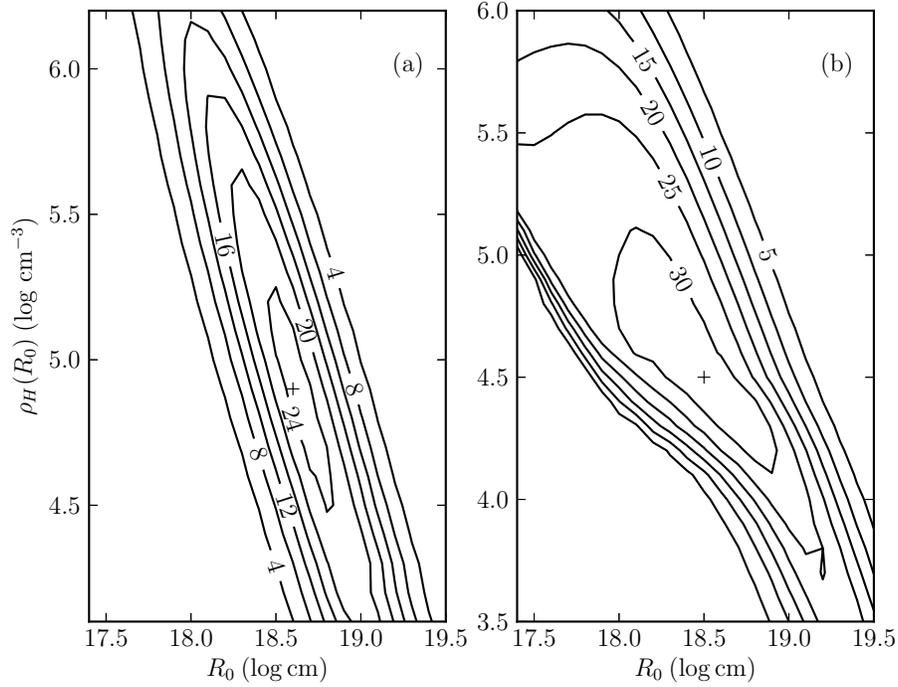}
    \caption{\ovh ratio contours in Cloudy model parameter space of a solar
      composition gas. The Cloudy models depend on three free parameters: inner
      radius of the gas distribution ($R_0$), hydrogen density at the inner
      radius ($\rho_H (R_0)$), and total gas mass. Panel (a) shows the \ovh
      ratio contours for 0.2 M$_{\sun}$ gas mass. At this mass the maximum ratio
      of 26 occurs at position indicated by the cross. A maximum ratio of
      \ovh=34 for all parameter space occurs for a gas mass of 100 M$_{\sun}$
      with contours for the other two free parameters shown in panel
      (b). }\label{fig:cloudy}
  \end{center}
\end{figure}

As an example of the resultant model space for one specific outflow mass,
Figure \ref{fig:cloudy}a displays the \ovh emission line ratios for a model
with a 0.2 $M_{\sun}$ outflow as a function of $R_0$ and $\rho_H(R_0)$.
Considering the case of a 0.2 $M_{\sun}$ outflow allows us to investigate an
interesting limit, since this is the minimum solar composition gas mass that
can provide the $\sim 10^{-4}$ $M_{\sun}$ of oxygen required to reproduce the
observed [O~III] emission \citep{Steele2011}.  For this 0.2 $M_{\sun}$ outflow
a maximum emission line ratio of 26 occurs at $R_0=4.0 \times 10^{18}$ cm and
$\rho_H(R_0)= 1.0 \times 10^5$ gm cm$^{-3}$.

The emission line ratio rises gradually with increased outflow mass reaching a
maximum of [O~III]$\lambda$5007/H$\beta = 34$ near $10^2 M_{\sun}$ as seen in
Figure \ref{fig:cloudy}b. The maximum emission line ratio for a range of
outflow masses is given in Figure \ref{fig:mvr}.  The transition point in the
maximum emission line ratio-mass ratio seen at a mass of $10^{34}$ g ($\sim 5$
$M_{\sun}$) occurs due to a change in the geometric distribution of the
emitting gas that maximizes \ovh . For masses below the $10^{34}$ g threshold
the optimal arrangement involves locating the gas in a thin shell ($R_0 /
\rm{Thickness} \sim 10^5$) at a density which allows maximum
   [O~III]$\lambda$5007 emission across the shell. Above the threshold mass the
   optimal gas configuration is of a thicker distribution ($R_0 /
   \rm{Thickness} \sim 10^2$) and the [O~III]$\lambda$5007 emission is
   generated over a broad region located primarily in the half of the gas
   distribution most distant from the ionizing source. For reference in a
   recent study \citet{Peacock12} find a half-light radius of 4-6 pc for the
   RZ2109 \zeroseven source, a few times greater than the outer radius
   producible by any solar composition model.

To evaluate the sensitivity of the [O~III]$\lambda$5007/H$\beta$ ratio on the
input X-ray source temperature, a second model grid was calculated employing a
higher temperature source. For an X-ray source temperature of $3 \times 10^6$ K,
the behavior of the maximum emission line ratio as a function of mass is largely
similar to the $1 \times 10^6$ K case described above. The transition point seen
produced by the shift in the optimal gas configuration occurs at a lower mass of
$3 \times 10^{33}$ g, and the maximum [O~III]$\lambda$5007/H$\beta$ ratio of
$\sim 40$ occurs near 30 $M_{\sun}$.  Over the range of masses evaluated the
higher temperature X-ray source produces line ratios 15--20 percent greater than
the corresponding lower temperature model.

The outflow velocity of 300 \kms was selected to be broadly consistent with the
observed 600 \kms feature, as the full observed width is produced by the
superposition of approaching and receding flows. Increasing the velocity to
1600 \kms to match the observed boarder feature produces a minimal change in
the resulting \ovh ratio; on order of a few thousandths of a percent.

\section{DISCUSSION AND CONCLUSION} \label{sec:dc}

From the EW measurements presented in Section \ref{sec:ew} it is clear that
any H$\beta$ emission which may be present in the observed RZ2109 spectrum is
extremely weak. The H$\beta$ emission is sufficiently weak to make
interpretation of the resulting \ovh line ratios difficult.  If taken at face
value these ratios are among the largest detected from any astrophysical source
\citep{Sarzi2006,Mendez2005}.  In Table \ref{tab:lim} the confidence levels
for the \ovh ratio are given based on the uncertainty in the H$\beta$
measurement.  Equivalent width measurements below the level of the continuum
(absorption features for example) are indicated with negative equivalent width
value. Therefore a negative line ratio limit in Table \ref{tab:lim} indicates
that the H$\beta$ measurement is consistent with an absorption line at the
specified confidence level.

When the full velocity width covered by the [O~III]$\lambda$5007 complex is
considered along with the radiative transfer modeling presented in Section
\ref{sec:rtm} it is clear that a solar composition gas is insufficient to
produce the observed \ovh ratios. The measured ratio using 3200 \kms apertures
is nearly a factor of three larger than the maximum ratios produced by the
synthetic emission line models.  At the $95\%$ confidence level for the 3200
\kms aperture and the $90\%$ confidence level for the 600 \kms aperture the
uncertainty limits on the \ovh ratio approach the maximum synthetic \ovh
values. These ratio limits are 35.7 at 3200 km~s$^{-1}$, and 33.0 at 600 \kms
compared to the synthetic maximum of 34.  As such it may be possible to produce
the necessary \ovh ratio for either aperture given an emission line region
model that falls in a very specific location is physical parameter space. It
should be noted, however, that the gas masses required to produce the maximum
synthetic \ovh values are well above what might be expected to be associated
with an accreting black hole in a globular cluster.  In order to produce the
maximum synthetic \ovh ratios a total gas mass of order 100 M$_{\sun}$ is
needed which would eliminate X-ray binaries, planetary nebulae, supernovae
remnants, or any other stellar scale objects as the source of the emitting gas.
To reach a gas mass of that size a significant portion of the gas would
necessarily be contributed by the cluster's interstellar medium. However the
hydrogen densities involved in producing the maximum synthetic \ovh ratios are
of order 10$^4$--10$^5$ cm$^{-3}$, orders of magnitude above the expected
densities of a cluster interstellar medium. Furthermore, the 10$^{-1}$ solar
composition of an interstellar medium would produce lower \ovh ratios in
comparison to the solar composition models considered above, making a large
contribution of ISM material to the emission line region unlikely.  More
typical gas masses produce maximum ratios a factor of four times smaller than
those observed in the 3200 \kms aperture and nearly twice that of the lower
limit of the 600 \kms measurement. The RZ2109 emission line region therefore
appears to be oxygen enriched relative to solar composition.

The level of oxygen enrichment necessary to produce the observed \ovh is not
readily apparent, as the emission line systems physical parameters are not well
constrained by observation. However, it is possible to examine some interesting
limits by calculating the affect of carbon and oxygen enrichment for two
characteristic masses; 1.0 M$_{\sun}$ representing a stellar source of the
emitting gas, and 0.1 M$_{\sun}$ representing the outflow from a CO white dwarf
and black hole binary.  A grid of cloudy models, as described in section 3.3,
was run for each characteristic mass with carbon and oxygen enriched from solar
composition to 100 times solar. From this calculation we find a minimum CO
enrichment of 17.8 times solar is necessary for the 1.0 M$_{\sun}$ model to
yield \ovh ratios equivalent to the measured values of the 3200 \kms aperture
and 5.5 times solar to match the 600 \kms aperture. For the 0.1 M$_{\sun}$
model the gas must be enriched to 17.2 times solar for the 3200 \kms aperture
and 5.6 times solar to match the 600 \kms aperture.  We must emphasize that
these are only lower limits on CO enrichment for two interesting astrophysical
systems, and are not necessarily indicative of enrichment of the RZ2109
emission line source.

\citet{Steele2011} present the argument that the complex line [O~III] velocity
profile observed in the RZ2109 spectrum is consistent with emission originating
from two discrete gas structures. The two velocity apertures presented in this
work do not directly correspond to the two geometric components discussed by
\citet{Steele2011} as emission from the two components are superimposed in
velocity space. From a comparison of the measured \ovh line ratios and synthetic
line ratio considered here it seems most likely that both the gas structures
are comprised of oxygen enriched material.  For the higher velocity structure
which \citet{Steele2011} describes as a bipolar conical outflow this is
consistent with the scenario of material being stripped from a CO white dwarf
companion to the black hole, and driven to the observed velocity as an
accretion-powered outflow. The added constraint of being oxygen enriched, does
not, however place obvious constraints on the gas source for the lower velocity
component.

\begin{figure}
  \begin{center}
    \includegraphics[scale=0.8]{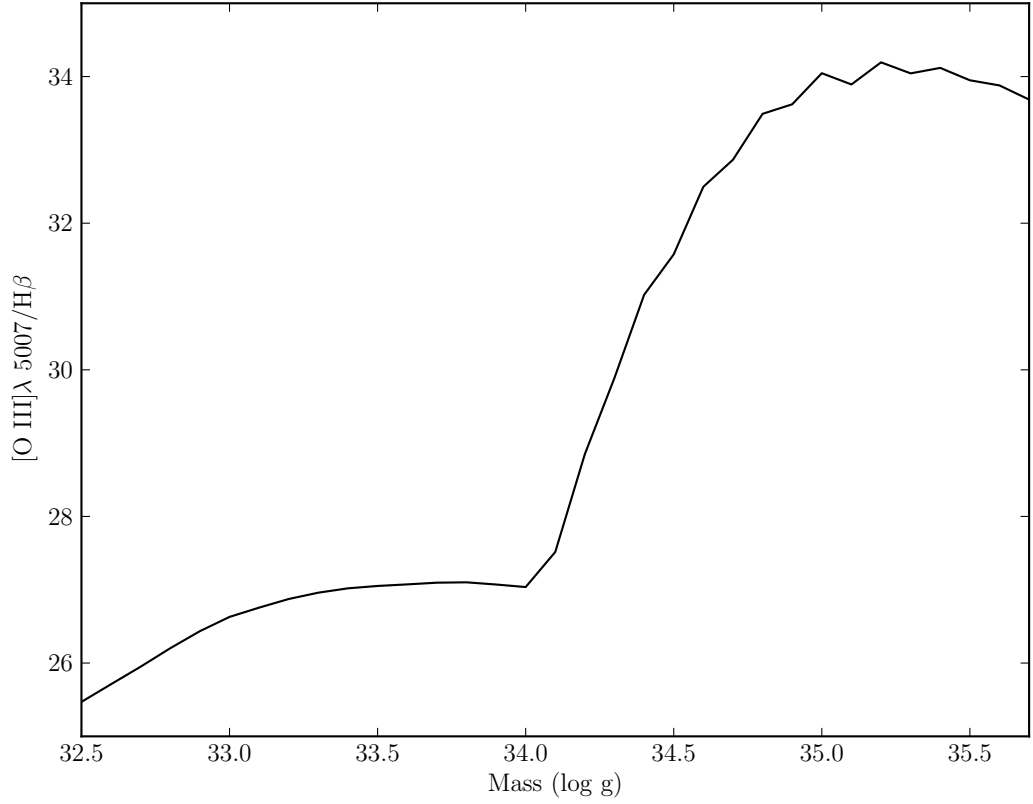}
    \caption{Maximum \ovh ratios by gas mass. The trend-line depicts the maximum
      \ovh ratio from Cloudy models for any given total gas mass, allowing the
      other free parameters (inner radius of the gas distribution and hydrogen
      density at the inner radius) to vary as necessary. The maximum ratio
      occurs with a total gas mass of $\sim 100$ M$_{\sun}$. }\label{fig:mvr}
  \end{center}
\end{figure}

As evidenced by the calculations summarized in Figure \ref{fig:mvr}, the
geometry of the emission region strongly influences the emission line ratios
that it produces.  As such the \ovh ratio alone is insufficient to fully
constrain the gas composition of the RZ2109 emission line system, or to
positively identify the particular stellar type of the X-ray binary's donor
star. The WD donor star model is consistent with all the observations and
calculations presented above. However we are not yet able to fully rule out
other late stage or polluted stellar types with super-solar abundance
atmospheres as the source for the observed outflow. With future observations of
other spectral bands, including measurements of UV carbon lines CIV 1548 \&
1551 \AA\ and CIII] 1907 \& 1910 \AA, and more precise estimates of the
  emission line system's gas mass, it may yet be possible to place tighter
  constraints on the emission line region's composition and the X-ray binary
  source system.

\acknowledgements M.M.S. and S.E.Z. wish to acknowledge support from NSF grant
AST-0807557. Support for A.K.’s work was provided by NASA through Chandra grant
numbers GO1-12112X and GO0-11111A issued by the Chandra X-ray Observatory.
K.L.R. acknowledges support from NSF Faculty Early Career Development grant
AST-0847109.  We thank the anonymous referee for the questions and critique
which helped to strengthen the paper. We also thank Jack Baldwin for helpful discussion.

\begin{deluxetable}{lcccc}
  \tablewidth{0pt}
  \tablecaption{Emission line equivalent widths and H$\beta$
    ratios \label{tab:ew}}
  \tablehead{
    \colhead{Species} & \colhead{Rest Wavelength (\AA)} &
    \colhead{Aperture (\kms)} & \colhead{EW (\AA)} &
    \colhead{X/H$\beta$}
  }
  \startdata

  H I & 4861 & 3200 & $0.32 \pm 0.32$ & 1.0 \\
  & & 600 & $0.21 \pm 0.13$\tablenotemark{a} & 1.0 \\
  {[}O III] & 5007 & 3200 & $33.82 \pm 0.39$ & 105.7 \\ 
  & & 600 & $13.14 \pm 0.10$ & 61.6 \\
  \enddata
  \tablenotetext{a}{The cited uncertainty reflects an 0.12 \AA\ observation
    measurement uncertainty and an 0.04 \AA\ stellar component model measurement
  uncertainty, which have been added by quadrature.}
\end{deluxetable}

\begin{deluxetable}{ccc}
  \tablewidth{0pt}
  \tablecaption{Cloudy Model Grid Parameter Space \label{tab:cmg}}
  \tablehead{\colhead{Mass Range (log g)} & \colhead{$R_0$ Range (log cm)}
    & \colhead{$\rho_H(R_0)$ (log cm$^{-3}$)}
  }

  \startdata
  31.1 -- 31.4 & 18.3 -- 19.0 & 4.2 -- 5.3 \\
  31.5 -- 33.5 & 17.8 -- 19.0 & 4.5 -- 5.5 \\
  33.6 -- 34.2 & 17.8 -- 19.0 & 4.4 -- 5.7 \\
  34.3 -- 35.3 & 17.8 -- 19.0 & 4.5 -- 5.5 \\
  35.4 -- 35.7 & 18.2 -- 19.0 & 4.0 -- 4.4 \\
  \enddata
  
\end{deluxetable}

\begin{deluxetable}{cccc}
  \tablewidth{0pt}
  \tablecaption{\ovh Ratio Limits \label{tab:lim}}
  \tablehead{
    \colhead{Confidence (\%)} & \colhead{Aperture
      (km s$^{-1}$)} & \colhead{Ratio, Upper H$\beta$ Limit} & \colhead{Ratio,
      Lower H$\beta$ Limit} 
  }
  \startdata

68.3 & 3200 &52.84 &   $\infty$ \\  
90.0 & 3200 & 39.96 &	-163.9 \\
95.0 & 3200 & 35.70 &	-110.1 \\
99.0 & 3200 & 29.55 &	-67.06 \\
68.3 & 600 &	41.05 &	165.4 \\
90.0 & 600 &	33.04 &	7186. \\
95.0 & 600 &	30.17 &	-364.3 \\
99.0 & 600 &	25.78 &	-119.3 \\

  \enddata
\end{deluxetable}

\end{document}